\documentclass[a4paper]{article}
\usepackage{Odyssey2020}
\usepackage{amsmath,amssymb,graphicx,booktabs,subfigure,caption}
\usepackage{url}

\ninept

\title{A Speaker Verification Backend for Improved Calibration Performance \\ across Varying Conditions}

\name{Luciana Ferrer$^1$, Mitchell McLaren$^2$}

\address{  $^1$Instituto de Investigaci\'on en Ciencias de la Computaci\'on (ICC), CONICET-UBA, Argentina\\
  $^2$Speech Technology and Research Lab (StarLab), SRI International, USA \\
\small \tt {lferrer@dc.uba.ar, mitchell.mclaren@sri.com}}

\DeclareMathOperator{\Norm}{Norm}
\DeclareMathOperator{\softmax}{softmax}

\begin{document}
\ninept

\setlength{\abovedisplayskip}{5pt}
\setlength{\belowdisplayskip}{5pt}

\maketitle

\begin{abstract}
In a recent work, we presented a discriminative backend for speaker verification that achieved good out-of-the-box calibration performance on most tested conditions containing varying levels of mismatch to the training conditions. This backend mimics the standard PLDA-based backend process used in most current speaker verification systems, including the calibration stage. All parameters of the backend are jointly trained to optimize the binary cross-entropy for the speaker verification task. Calibration robustness is achieved by making the parameters of the calibration stage a function of vectors representing the conditions of the signal, which are extracted using a model trained to predict condition labels. In this work, we propose a simplified version of this backend where the vectors used to compute the calibration parameters are estimated within the backend, without the need for a condition prediction model. We show that this simplified method provides similar performance to the previously proposed method while being simpler to implement, and having less requirements on the training data. Further, we provide an analysis of different aspects of the method including the effect of initialization, the nature of the vectors used to compute the calibration parameters, and the effect that the random seed and the number of training epochs has on performance. We also compare the proposed method with the trial-based calibration (TBC) method that, to our knowledge, was the state-of-the-art for achieving good calibration across varying conditions. We show that the proposed method outperforms TBC while also being several orders of magnitude faster to run, comparable to the standard PLDA baseline.
\end{abstract}
\section{Introduction}
\label{sec:intro}
Most current speaker verification systems are composed of several separate stages. First, frame-level features that represent the short-time contents of the signal are extracted. These features are input to a deep neural network (DNN) which is trained to optimize speaker classification performance on the training dataset. A hidden layer within that DNN is then used as a signal-level feature extractor. These new features, termed speaker embeddings or `x-vectors'~\cite{snyder2016deep}, are transformed using linear discriminant analysis (LDA), then mean- and variance-normalized and, finally, length normalized. Next, probabilistic linear discriminant analysis (PLDA) is used to obtain scores for each speaker verification trial. Finally, a calibration stage is necessary to convert the scores produced by PLDA into proper log-likelihood ratios (LLRs) that can be thresholded to make decisions or used directly. This stage usually consists of an affine transformation of the scores where parameters are trained to optimize a weighted binary cross-entropy objective which measures the overall quality of the scores as proper LLRs. 
Assuming the calibration training data reflects the evaluation conditions, this procedure has been repeatedly shown to provide hard-to-beat performance on a wide range of datasets. 

In a recent paper \cite{Ferrer:dplda19}, we proposed an alternative backend which integrates all the steps from LDA to calibration into a single jointly-trained model. The functional form of this backend coincides with that of the standard backend described above, except for the calibration stage. In this final stage, instead of using a single set of trainable calibration parameters for all trials, these parameters are a function of vectors representing the conditions for the two sides of the speaker verification trial. These vectors are extracted from a layer in a DNN trained to predict the conditions present in a large set of training data. The transformation from condition vectors to calibration parameters is trained jointly with the rest of the model. In \cite{Ferrer:dplda19}, we show that the discrimination performance of the proposed model is comparable to that of the standard backend, while the calibration performance is, in most cases, as good as or better than the best of three possible calibration models trained with different subsets of data. Overall, the proposed model achieves a very low calibration loss in most test sets. 

This method, proposed in~\cite{Ferrer:dplda19} and extended in this paper, is related to recent papers~\cite{snyder2016deep,Rohdin2018} which also propose to use the binary cross-entropy as loss function during DNN training. In~\cite{snyder2016deep}, a DNN is used to obtain an embedding for each signal. The score for a trial is then computed using a simple function of the two embeddings with the same form as the PLDA scores, an idea that was first proposed in \cite{burget:icassp11}. The parameters for the embedding extractor and the scorer are trained jointly to optimize binary cross-entropy. In \cite{Rohdin2018}, the authors propose to use an architecture that mimics the previous i-vector \cite{Dehak11} pipeline for speaker verification, pre-training all parameters separately and then fine-tuning the full model to minimize binary cross-entropy. Neither of these papers show overall system performance
(i.e., including the effect of calibration),
only discrimination performance. In fact, as we show in our previous paper \cite{Ferrer:dplda19},  discriminatively training a PLDA-like backend as done in those papers does not suffice to achieve good generalization in terms of calibration. For this reason, we proposed to integrate the condition-aware calibration stage inside the backend, which significantly reduced calibration problems on most tested datasets. In the rest of this paper, we will call the method proposed in our previous paper condition-aware discriminative PLDA (CA-DPLDA). 

Several approaches have been proposed in the speaker verification literature which take into account the signal's conditions in different ways during calibration. In some cases, the side-information was assumed to be discrete and known (or estimated separately) during testing \cite{FerrerEtAl:Eurospeech2005,Solewicz:Eurospeech2005,solewicz:ASLP07,FerrerEtAl:icassp2008}, and calibration parameters were conditioned on these discrete side-information values. The Focal Bilinear toolkit \cite{FocalBilinear} implements a version of side-information-dependent calibration where the calibrated score is a bilinear function of the scores and the side-information vector, which is assumed to be composed of numbers between 0 and 1. More recently, we proposed an approach called trial-based calibration (TBC) where calibration parameters are trained independently for each trial using a subset of the development data \cite{mclaren2014trial,Ferrer:aslp18} selected using a model trained to estimate the similarity between the conditions of two samples. This approach, while successful, is quite computationally expensive and requires tuning a few different parameters in order to obtain good performance. In our proposed model, both discrete (in the form of one-hot vectors) and continuous side-information can be used and the functional form is a generalization of all previous approaches, except for TBC. In addition, in our proposed approach, the calibration model is trained jointly with the rest of the backend parameters, while in all previous approaches the calibration step was trained separately. 

The CA-DPLDA method requires a separate model, a DNN, trained to predict condition labels and used to extract condition vectors as the activations from a layer within the DNN. This requires training data labeled with information about the conditions present in each signal. While some datasets have this information available, others do not. In our work, we used as much information about the signals' conditions as was available in each of the sets used for training, which seemed to suffice. Yet, the need for a separate model has other disadvantages: the performance of the backend depends on how this model is trained, which data and labels are used to train it, which architecture is chosen and which seeds are used for initialization. All of these hyperparameters can be optimized for the final speaker verification performance after training the CA-DPLDA backend, but this is a slow and involved development process. 
For these reasons, we endeavored to eliminate the need for a condition-prediction model. 

In this paper we propose a simplified version of the CA-DPLDA approach, which we call automatic side-information DPLDA (AS-DPLDA). In this approach, the vectors used to obtain the calibration parameters are learned along with the rest of the backend as a function of the embeddings, without the need for condition labels. We call these vectors side-information vectors rather than condition vectors since they do not necessarily reflect the conditions present in the signals, given that they are not trained with this goal but only with the goal of optimizing speaker verification performance (i.e., to minimize cross-entropy for the binary speaker verification task).

The contributions of the current paper are as follows: (1) we propose a simplified version of the method introduced in \cite{Ferrer:dplda19} which does not require externally-computed condition vectors and show that it performs on par with the original method; (2) we compare the proposed method's performance with that of TBC \cite{Ferrer:aslp18} and show that our current method outperforms TBC, while also being orders of magnitude faster to run;  (3) we provide an analysis of the vectors used to compute calibration parameters both for the CA-DPLDA and the AS-DPLDA methods; and (4) we show the effect that the initialization method and seed and the number of epochs have on the performance of the method. In summary, this paper is an extension of our previous paper \cite{Ferrer:dplda19}, proposing a simplified version of the method introduced in that paper and showing a more detailed analysis of different aspects of the method.

\section{Standard PLDA-based Backend}
Most state of the art speaker verification systems consist of an embedding extraction stage followed by a PLDA-based backend. The PLDA-based backend is in itself composed of several stages. First, linear discriminant analysis is applied to reduce the dimension of the embeddings while emphasizing speaker information and reducing other irrelevant information. Then, each transformed dimension is mean- and variance-normalized (MVN) and the resulting vectors are length normalized. Finally, PLDA is used to compute a score for each trial. While the training procedure for PLDA is somewhat involved and requires the use of an expectation-maximization algorithm, once parameters have been trained, scoring is done with a simple function of the two embeddings involved in the trial (see \cite{cumani2013pairwise} for a derivation). 

To summarize, the set of equations required to go from two individual embeddings, $x_1$ and $x_2$, to a score $s$ for the trial are:
\begin{eqnarray}
\tilde x_i \hspace{-0.2cm} & = & \hspace{-0.2cm}\Norm( Px_i+\mu), \forall i \in \{1,2\}, \label{eq:lda} \\
s  \hspace{-0.2cm}& = & \hspace{-0.2cm}2 \tilde x_1 ^T \Lambda \tilde x_2 + \tilde x_1^T \Gamma \tilde x_1 + \tilde x_2^T \Gamma \tilde x_2 + \tilde x_1^Tc + \tilde x_2^T c + k,  \label{eq:plda_score}
\end{eqnarray}
where $P$ is the LDA projection matrix restricted to the first N dimensions and scaled to result in variance of 1.0 in each dimension, $\mu$ is the global mean of the data after multiplication with $P$, $\Norm$ performs length normalization, and $\Lambda$, $\Gamma$, $c$ and $k$ are derived from the parameters of the PLDA model using Equations (14) and (16) in \cite{cumani2013pairwise}. 

PLDA scores are given by the logarithm of the ratio between the likelihood for the hypothesis that the speakers in the two signals in the trial are the same and the likelihood of the hypothesis that the speakers are different (Equation \ref{eq:plda_score} computes this value, given the PLDA model). That is, the score is defined as a log-likelihood ratio (LLR). Yet, in practice, the scores produced by PLDA are far from being proper LLRs, i.e., they do not reflect the distributions found during evaluation: a PLDA score of log(2.0) cannot be interpreted as indicating that the likelihood of the same-speaker hypothesis is two times higher than that of the different-speaker hypothesis. This is due to the fact that PLDA's assumptions do not exactly hold in practice. Hence, the LLRs produced by the model are not well-calibrated. 
It is possible to use the raw scores from PLDA to make speaker verification decisions by tuning a threshold on some development data for the specific application of interest. Yet, in many cases, like in forensic applications, when no development data is available to choose a threshold, or 
when the operating point is not defined a priori, it is necessary for the system to output proper LLRs which can then be thresholded using Bayes rule for any cost of interest or directly used as  stand-alone interpretable values. To achieve this, an additional stage of calibration is needed.

The standard procedure for calibration in speaker verification is to use linear logistic regression, which applies an affine transformation to the scores, training the parameters to minimize binary cross-entropy \cite{brummer2013likelihood}.  The objective function to be minimized is given by
\begin{equation}
C_\pi = -\frac{\pi}{T} \sum_{k \in \cal T}  \log(q_k) - \frac{1-\pi}{N} \sum_{k \in \cal N}  \log(1-q_k), \label{eq:crossent}
\end{equation}
where
\vspace{-0.2cm}
\begin{eqnarray}
q_k & = &\sigma\left(l_k + \log (\pi /(1-\pi))\right), \\
l_k & = & \alpha s_k + \beta, \label{eq:cal}
\end{eqnarray}
and where $s_k$ is the score for  trial $k$ given by Equation (\ref{eq:plda_score}), $\sigma$ is the sigmoid function, $\pi$ is a parameter reflecting the expected prior probability for same-speaker trials, and $\alpha$ and $\beta$ are the calibration parameters, trained to minimize the quantity in Equation~(\ref{eq:crossent}).

The top part of Figure \ref{fig:architecture} (the orange blocks) show the stages in the standard PLDA pipeline, referencing the equation implemented in each stage. The parameters involved in these equations are all trained separately, freezing the parameters of the previous steps in order to obtain input data to train the next step. 

\section{Discriminative Backend with Side-Information-Dependent Calibration}
\label{sec:dplda}

In a recent paper \cite{Ferrer:dplda19} we present a backend method with the same functional form as the PLDA-backend explained in the previous section, but where all parameters are optimized jointly, in a manner similar to the one used in \cite{Rohdin2018} (though, note that in this paper we only optimize jointly up to the backend stage instead of the full pipeline, as in Rohdin's paper). 
In this method, we first initialize all parameters in Equations (\ref{eq:lda}), (\ref{eq:plda_score}) and (\ref{eq:cal}) as in the standard PLDA-based backend. Then, we fine tune the parameters to optimize the cross-entropy using Adam optimization~\cite{kingma2014adam}. To this end, we define mini-batches that contain both same-speaker and different-speaker samples. This is done by randomly selecting N speakers for each mini-batch. Then, two random samples from each of those speakers are chosen. All possible trials between the 2N selected samples are used to compute the cross-entropy, after excluding all same-session target trials and different-domain impostor trials. We found that these two restrictions were important to get good calibration performance. 

In \cite{Ferrer:dplda19}  we show that having a global calibration model with two trainable parameters $\alpha$ and $\beta$ as in Equation (\ref{eq:cal}) does not suffice to get good generalization in terms of calibration performance.  This problem can be fixed, as usually done for the standard PLDA backend, by training a specific calibration model for each domain of interest, which requires having at least some domain-specific labeled data. 
In this research, though, we assume that no domain-specific data is available for system adaptation or for training a calibration model. This also means that a domain-specific decision threshold cannot be learned. Hence, we aim to design the best possible out-of-the-box system for unknown conditions for which the  score produced can be thresholded using the theoretically optimal threshold assuming the scores are proper LLRs (see, for example, Equation (6) in \cite{van2007introduction}). 

In order to achieve this goal, in \cite{Ferrer:dplda19}, we proposed to make the calibration parameters depend on the conditions of the signal by having the calibration scale and shift, $\alpha$ and $\beta$ in Equation (\ref{eq:cal}), be functions of side-information vectors, $z_1$ and $z_2$, for each of the signals in a trial:
\begin{eqnarray}
\alpha  =  2 z_1^T \Lambda_\alpha z_2 + z_1^T \Gamma_\alpha z_1+ z_2^T \Gamma_\alpha z_2 + (z_1 + z_2)^T c_\alpha + k_\alpha, \label{eq:alpha} \\
\beta   =  2 z_1^T \Lambda_\beta z_2 + z_1^T \Gamma_\beta z_1 + z_2^T \Gamma_\beta z_2 + (z_1 + z_2)^T c_\beta + k_\beta. \label{eq:beta} 
\end{eqnarray}
In our implementation, all parameters in these equations are initialized to 0 except the $k$ values that are initialized using the  global calibration parameters trained using linear logistic regression. Hence, at initialization, the calibration stage coincides with the global calibration model.

The key component of this model are the vectors $z_i$, which we define to be given by
\begin{equation}
z_i = \log \softmax(W m_i), \label{eq:meta}
\end{equation}
where $m_i$ was, in our original proposed method, an additional input to the backend extracted from a bottleneck layer from a DNN trained to predict the conditions in the signal. Several other options were tested to transform $m_i$ into $z_i$: adding a bias terms, using length-normalization, no transformation, softmax without the logarithm and relu. None of these alternatives proved to be better in our experiments than the logarithm of the softmax transformation. Also, the trainable transformation proved to be essential to obtain good performance. Using the pre-activations (or activations) from the condition DNN directly as $z$ values with or without log softmax led to suboptimal results. In our experiments the $W$ parameter is trained jointly with the rest of the model. This is the only parameter that is initialized randomly using a normal distribution centered at 0.0 with standard deviation of 0.5.

\begin{figure}[!t]
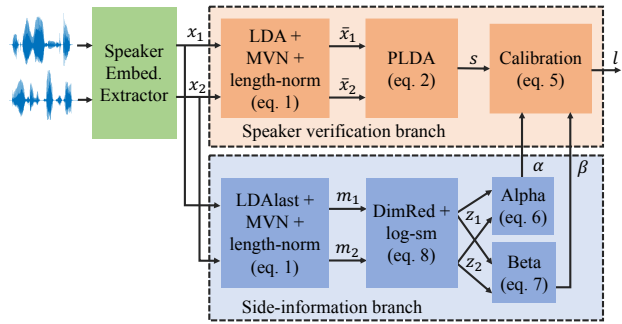

\centering
\includegraphics[width=1.02\columnwidth]{{{figs/DPLDA_backend}}}
\vspace{-0.6cm}
\caption{Schematic of the baseline PLDA and the proposed AS-DPLDA backends. The orange blocks at the top correspond to the standard PLDA pipeline. The proposed AS-DPLDA includes the same components as the baseline system plus the lower blocks which extract the calibration parameters as a function of the same speaker embeddings. The equations indicated in parenthesis inside each block describe the functional form of the block. For the AS-DPLDA backend the name of the block indicates how the parameters are initialized before proceeding to jointly training the whole backend. } 
\label{fig:architecture}
\end{figure}

In the current work, we propose a simpler alternative to extract the $m_i$ vectors that does not require training a separate condition-prediction DNN and, hence, does not require training data with condition labels. In this alternative, the $m$ vectors are given by an affine transform of the embeddings followed by length normalization. This affine transform is initialized using the last M dimensions of the same LDA transform used to initialize the speaker verification branch of the model (Figure \ref{fig:architecture}), including mean and variance normalization. That is, we use the dimensions that are less useful for speaker verification under the LDA assumptions and apply the same procedure as the LDA stage in the speaker verification branch. After initialization, this transform, which has the form in Equation \ref{eq:lda}, is fine-tuned along with all the other parameters in the model.

Figure \ref{fig:architecture} shows the complete architecture of the proposed backend. The names in the blocks refer to the way the parameters of that stage are initialized. Without the lower blue part, the proposed backend coincides, at initialization, with the standard PLDA backend with a global calibration stage. LDAlast refers to using the last M dimensions of the LDA matrix for initialization of that block. As mentioned before, the parameters of the dimensionality reduction applied to the $m$ vectors to obtain the $z$ vectors (the DimRed block) are the only ones that are randomly initialized. Note that the blocks Alpha and Beta are in charge of computing the calibration parameters when those are dependent on the side-information. That is, in this case, the parameters become an input to the calibration stage while, when no side-information is used, the $\alpha$ and $\beta$ parameters are global and are implicit in the calibration block. 

\section{Trial-Based Calibration}

Through the years we have been exploring the problem of achieving robust calibration performance across varying conditions using different approaches. The most successful of them, published first in \cite{mclaren2014trial} and then further developed and analyzed in \cite{Ferrer:aslp18} was trial-based calibration (TBC). The approach consists of training a separate calibration model for each verification trial using a subset of the available calibration data. The subset for each trial is selected based on a measure of condition similarity between the two sides of the trial and the calibration data. That is, for each trial we select calibration enrollment samples that are similar in terms of condition to the enrollment side of the trial to be scored, and calibration test samples that are similar to the test side of the trial to be scored. Next, we train a calibration model using all possible trials generated with the selected enrollment and test samples. In the latest version of the method \cite{Ferrer:aslp18}, the similarity between two signals was measured using a PLDA model trained with condition rather than speaker labels. In this paper, we use the parameters we found to be best in \cite{Ferrer:aslp18}, under the assumption that we want to score every possible trial. A reject option was proposed in that paper, but we do not use it for the comparisons here since we haven't implemented an equivalent approach for DPLDA yet.

\section{Experimental Setup}
In this section we describe the system configuration and datasets used for our experiments. 

\subsection{Speaker Recognition System}
\label{sec:system}
The proposed backend uses standard x-vectors as input \cite{KaldiRecipe17}. 
The input features for the embedding extraction network are power-normalized cepstral coefficients (PNCC)~\cite{kim:12} which, in our experiments, gave better results than the more standard mel frequency cepstral coefficients (MFCCs). We extract 30 PNCCs with a bandwidth going from 100 to 7600 Hz and root compression of 1/15. The features are mean and variance normalized over a rolling window of 3 seconds. Silence frames are then discarded using a DNN-based speech activity detection system.

System training data included 234K signals from 14,630 speakers. This data was compiled from NIST SRE 2004--2008, NIST SRE 2012, Mixer6, Voxceleb1, and Voxceleb2 (train set) data. Voxceleb1 data had 60 speakers removed that overlapped with Speakers in the Wild (SITW). All waveforms were up- or down-sampled to 16 KHz before further processing.  In addition, we down-sampled any data originally of 16 kHz or higher sampling rate (74K files) to 8 kHz before up-sampling back to 16 kHz, keeping two ``raw'' versions of each of these waveforms. This procedure allowed the embeddings system to operate well in both 8kHz and 16kHz bandwidths.

Augmentation of data was applied using four categories of degradations as in~\cite{mclaren:odyssey18}, including music and noise, both at 10 to 25 dB signal-to-noise ratio, compression, and low levels of reverb. We used 412 noises compiled from both freesound.org and the MUSAN corpus. Music degradations were sourced from 645 files from MUSAN and 99 instrumental pieces purchased from Amazon music. For reverberation, examples were collected from 47 real impulse responses available on echothief.com and 400 low-level reverb signals sourced from MUSAN. Compression was applied using 32 different codec-bitrate combinations with open source tools. We augmented the raw training data to produce 2 copies per file per degradation type (randomly selecting the specific degradation and SNR level, when appropriate) such that the data available for training was 9-fold the amount of raw samples. In total, this resulted in 2,778K files for training the speaker embedding DNNs. 

The architecture of our embeddings extractor DNN follows the Kaldi recipe~\cite{KaldiRecipe17}. The DNN is implemented in Tensorflow, trained using an Adam optimizer with chunks of speech between 2.0 and 3.5 seconds. Overall, we extract about 4K chunks of speech from each of the speakers. 
DNNs were trained over 4 epochs over the data using a mini batch size of 96 examples. 
We used dropout with a probability linearly increasing from 0.0 up to 0.1 at 1.5 epochs then linearly decreasing back to 0.0 at the final iteration. The learning rate started at 0.0005, increasing linearly  after 0.3 epochs reaching 0.03 at the final iteration while training simultaneously using 8 GPUs, averaging the parameters from the 8 jobs every 100 mini-batches.

The training data for the PLDA and DPLDA backends was a subset of the training data used for the speaker embeddings DNN including a random half of the speakers (for expedience of experimentation) and excluding all signals for which no information about the recording session could be obtained and all speakers for which a single session was available.  In this case, we use full segments to train the backend rather than chunks and SNR level of 5dB for augmentation (using this SNR on the PLDA backend led to marginally better results than using 10-25dB SNR, as for the embeddings extractor). 
Beside this training data, we add two datasets for backend training, FVCAUS and RATS.  FVCAUS is composed of interviews and conversational excerpts from over 500 Australian English speakers from the forensic voice comparison dataset~\cite{morrison2015forensic}. Audio was recorded using close talking microphones. RATS is composed of telephone calls in five non-English languages from over 300 speakers. We only used the source data (not retransmitted) of the DARPA RATS program~\cite{ref:rats_set} for the SID task. 


Both for PLDA and CA-DPLDA, the LDA dimension was found to be optimal at 200. For AS-DPLDA the optimal LDA dimension turned out to be 300. The effect of changing this dimension from 200 to 300 is small for all three systems. 

The condition DNN used to generate the embeddings $m_i$ for the CA-DPLDA method has two layers of 100 and 10 nodes with ReLU activations and batch normalization. The classes used at the output layer are given by the domain (Voxceleb, Mixer, Switchboard, FVCAUS or RATS) concatenated with the degradation type and, when available, any further information about the condition of the signal (channel type, language, and speech style). Note that the classes are then extremely different in terms of granularity. All Voxceleb data is grouped into one class per degradation type, while Mixer data has much finer grained labels. While this is clearly suboptimal, it seems to work well in our experiments. For TBC, we train a condition-PLDA model using the same data and labels as for the condition DNN above. In the case of the AS-DPLDA method, we do not need a separate condition prediction model. We simply initialize the transformation from embeddings to $m$ vectors using the last 200 dimensions in the LDA matrix
out of the 512 dimensions available.

\subsection{Two-stage training}
\label{sec:two-stage}
Our backend training data, described in Section \ref{sec:system}, is highly imbalanced: 53\% comes from Voxceleb collections, 25\% from SRE and Mixer collections, 11\% from Switchboard, 6\% from RATS, and 4\% from FVCAUS. This causes a problem when learning the calibration part of the model, since parameters cannot be robustly learned for the underrepresented conditions. For this reason, we implement a two-stage training procedure. We use all the training data for the first few iterations, then freeze the parameters of the speaker verification branch up the score generation stage (Equation \ref{eq:plda_score}), subset the training list to use a balanced set of samples with similar representation for all five domains and continue training the calibration parameters. This allows the model to focus on improving side-information-dependent calibration once the discriminative part of the model has converged. 

\subsection{Datasets}
\label{sec:evaldata}

We use several different datasets for development and evaluation of the proposed approach. Table \ref{tab:dsets} shows the statistics for all sets.
The {\bf SITW} dataset contains speech samples in English from open-source media~\cite{sitwdb} including naturally occurring noises; reverberation; codec; and channel variability. 
The {\bf SRE16} dataset \cite{sre16} includes variability due to domain/channel and language mismatches. 
We use the CMN2 subset of the {\bf SRE18} dataset \cite{sre18}, which has similar characteristics to the SRE16 dataset, with the exception of focusing on different languages, and including speech recorded over VOIP instead of just PSTN calls.
The {\bf LASRS} corpus is composed of 100 bilingual speakers from each of three languages, Arabic, Korean and Spanish~\cite{Beck2004}. Each speaker is recorded in two separate sessions speaking English and their native language using several recording devices.  
Finally, the {\bf FVCCMN} is composed of interviews and conversational excerpts from female Chinese speakers~\cite{zhang2011forensic}, which were cut to durations between 10 and 60 seconds. Recordings were made with high-quality lapel microphones.

\begin{table}[!t]
\centering
\vspace{0.1cm}
\caption{The development and evaluation datasets with number of speakers (spk) and target/impostor (tgt/imp) trial counts.}
\vspace{-0.1cm}
\footnotesize
\begin{tabular}{l|rrr|rrr}
	Dataset & \multicolumn{3}{c|}{Dev Split} &  \multicolumn{3}{c}{Eval Split} \\
	              & \#spk & \#tgt & \#imp & \#spk & \#tgt & \#imp  \\ 
	\midrule
	SITW  &  119 & 2.6k & 335.0k &  180 & 3.7k &  0.7m \\
	SRE16  &  20 & 3.3k & 13.2k &  201 & 27.8k & 1.4m \\
	SRE18  &  35 & 6.3k & 80.2k  &  289 & 48.5k & 1.6m \\
	FVCCMN & - & - & - & 68 & 16.4k & 1.1m \\
	LASRS & - & - & - &  333 & 41.0k & 4.8m \\
	\bottomrule
\end{tabular}
\label{tab:dsets}
\end{table}

%
%
%
%

SITW, SRE16 and SRE18 have well-defined development sets. We use those 3 sets to tune the parameters of our models. The rest of the sets are used for evaluation of the final systems. For SITW, SRE16 and SRE18 we use the 1-side enrollment trials defined with the datasets. For LASRS and FVCCMN we create exhaustive trials excluding same-session trials.

\section{Results}

We show results in terms of Cllr. This metric \cite{Brummer:csl06} measures the quality of the scores as LLRs using a logarithmic cost function and is affected both by the discrimination and calibration performance of the system. A very discriminant system can have a high Cllr if the calibration is wrong (ie, if the scores do not represent proper LLRs for the task). Such a system would lead to bad decisions when thresholded with the theoretically optimal threshold for the cost function of interest. In this work, we aim to obtain a system that results in good calibration across a large variety of conditions. To measure whether we are succeeding in this goal, we need to separate the effect of the discrimination and the calibration performance of the system. This is done by obtaining the minimum Cllr that can be achieved with the system's scores for a certain test set using a monotonic transformation \cite{brummer:pav}. The difference between the actual Cllr and minimum Cllr for the system indicates the effect of miss-calibration. If the two values are equal, then the system is perfectly calibrated and the scores produced by it are proper LLRs.

\begin{figure*}[!t]
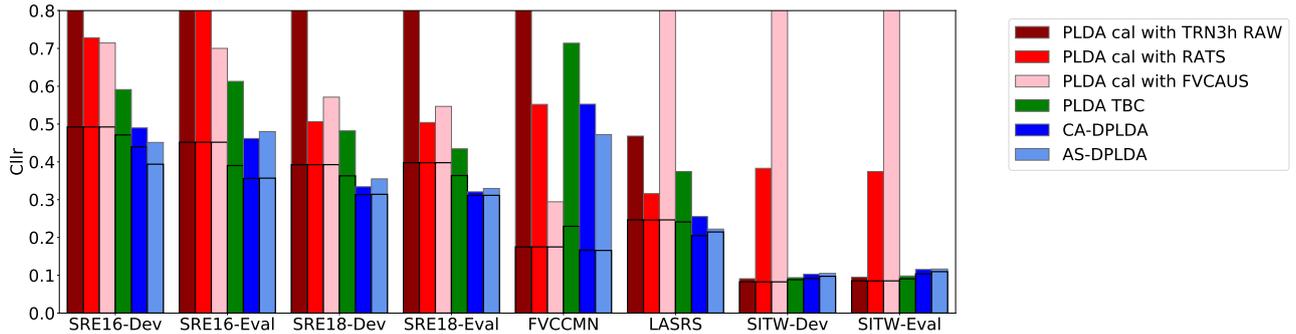

\centering
\includegraphics[width=1.0\textwidth]{{{figs/TABLE.plda_vs_dplda.ACLLR}}}
\vspace{-0.6cm}
\caption{Actual CLLR (bar height) and minimum Cllr (black line inside bars) for different PLDA and DPLDA systems on all test sets. Bars taller than 0.8 are cut to allow better resolution of all other results. } 
\label{fig:res}
\end{figure*}

Figure \ref{fig:res} shows the actual and minimum Cllr values for development and evaluation dataset for different systems. All development decisions (dimensions, training hyperparameters, number of epochs, initialization seed, etc) were made based on the average performance in all three development sets. During our initial experiments for \cite{Ferrer:dplda19} we concluded that including the $\Gamma_\alpha$ and $\Gamma_\beta$ terms in Equations (\ref{eq:alpha}) and (\ref{eq:beta}) did not improve results. Hence, those terms are not used in our experiments. The results shown correspond to the best epoch of each of 10 models run with different seeds for the chosen architecture, selected based on the average development set performance, since, as we will see in Section \ref{sec:init}, the seed and the number of epochs have a very significant effect in system performance. 

For completeness sake, we repeat here the baseline PLDA results shown in \cite{Ferrer:dplda19}. 
For these three PLDA systems the LDA and PLDA parameters are trained using only the training data, without adding FVCAUS and RATS since adding those sets slightly degrades discrimination performance of this system on our development sets. We show three options for training the calibration stage for the PLDA system: using TRN3h RAW which consists of 300 speakers from the raw part of the training set (using more speakers does not help and including the degraded part hurts performance), using only RATS data and using only FVCAUS data. Note that no calibration set is optimal for all test sets. Merging the three sets leads to a trade-off in performance which highly depends on the proportion of each dataset used (results not shown). Note also that the discrimination performance of the baseline system is not affected by the calibration model, since this model is a single monotonic transformation for each test set.

The fourth system in the figure uses the same PLDA backend as the first three and TBC for condition-dependent calibration. In this case, the concatenation of the three calibration sets used for the first three systems is given to the TBC algorithm as candidate set for selection of calibration data adding back the degraded training data, since including that data provides improved performance on the SITW-Dev set and no degradation on the other two development sets. Ideally, the TBC algorithm should be able to select the best subset of data for each trial. We set TBC to select enough samples to achieve 100 target trials, and to use regularization toward the global calibration parameters, with a weight of 0.02. These parameters were found to give optimal results in \cite{Ferrer:aslp18}. As we can see, TBC is working as expected, giving, in most cases, a performance better than or similar to each of the three global calibration models, with a clear exception for FVCCMN (more on this issue below).

Finally, we show results for two DPLDA systems. The first system is the one proposed in \cite{Ferrer:dplda19} and corresponds to the system called ``DPLDA with meta 2stage'' in that paper. Results are slightly different from those in the paper since we used the best of 10 seeds here instead of 5 as in the original paper. The last system in the figure is the newly proposed variant, AS-DPLDA, where no external condition prediction model is used to extract the $m$ vectors. As we can see, both DPLDA methods give very similar performance, both in terms of discrimination and calibration, with the newly proposed method being much simpler. 

We can see that the only data set that remains significantly misscalibrated with the DPLDA methods is FVCCMN.  This dataset is quite different from all our training data. While it is similar in terms of acoustic conditions to FVCAUS, it consists only of Chinese speech while FVCAUS consists of English speech with Australian accent. While around 1\% of the training samples are in Chinese (all from SRE04 and SRE06 datasets), they are all recorded over a telephone channel, not in the extremely clean acoustic conditions of the FVCCMN dataset. Note that the FVCAUS dataset leads to reasonable calibration when used as training data for calibrating FVCCMN (pink bar) despite the fact that the languages in these two sets are different. Our DPLDA algorithms, on the other hand, do not believe that the FVCAUS data should be relevant to calibrate FVCCMN given this difference in language. This is probably the reason why the DPLDA (and TBC) methods give worse performance than PLDA with FVCAUS calibration.

In terms of run time, the PLDA baseline with global calibration and the DPLDA options take a similar amount of time to run since, once the model is trained, both have the same functional form, with the DPLDA system having a slight increase in run time due to the computation of the $z$ vectors. On the other hand, the TBC approach takes orders of magnitude more time than either PLDA with global calibration or DPLDA. For example, the evaluation for the SITW-Eval set took about 3000 minutes on a single CPU for the PLDA system with TBC, while the PLDA with global calibration and DPLDA systems took under 10 seconds on a single CPU.

\subsection{Analysis of the side-information vectors}
\label{sec:metadata}
Using the $z$ vectors as input to a simple functional form (Equations \ref{eq:alpha} and \ref{eq:beta}) to extract the calibration parameters appears to be quite successful at achieving robustness across different datasets. So, a question arises: what information do these $z$ vectors represent? For the original model proposed in \cite{Ferrer:dplda19}, where the $m$ vectors are given by the pre-activations in a 10-node bottleneck layer in a condition-prediction DNN, we expect the resulting $z$ vectors to contain mostly condition information and very little speaker information. On the other hand, for the AS-DPLDA method, the $z$ vectors could, in fact, contain speaker information, since they are not prevented otherwise.

\begin{figure}
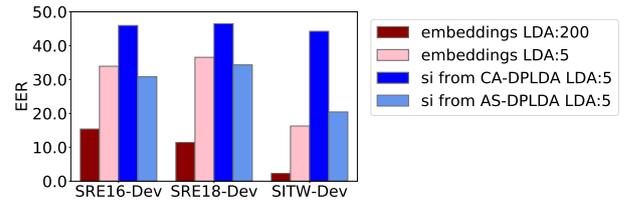

\centering
\includegraphics[width=1.02\linewidth]{{{figs/TABLE.sid_from_meta.EER}}}
\vspace{-0.6cm}
\caption{EER for different PLDA systems using different LDA dimensions and different input features: embeddings or the side-information (si)  vectors extracted from the DPLDA methods.}
\label{fig:sid_from_meta}
\end{figure}

Figure \ref{fig:sid_from_meta} shows system performance when using the $z$ vectors from the two DPLDA systems in Figure \ref{fig:res} as input to a standard PLDA-based speaker verification system. In this case, the input is 5-dimensional, so, the LDA transform is set to not reduce dimensionality further. We choose to show EER in this case since we do not care about calibration performance, we only wish to assess how much speaker information is present in the $z$ vectors from each system. For comparison, we also show the performance for two PLDA systems using the speaker embeddings as input, one with LDA dimension of 200 (this is the same system used in Figure \ref{fig:res} for the first three results) and one with LDA dimension of 5. 


As we can see, using only 5 dimensions (obtained from the speaker embeddings in different ways) is significantly worse than using 200; no surprise there. Interestingly, though, using the AS-DPLDA $z$ vectors gives a similar performance to the PLDA system with embeddings as input and an LDA dimension of 5. Hence, we need to assume that this system is probably prioritizing the preservation of speaker information within those 5 dimensions. On the other hand, the PLDA system using the CA-PLDA $z$ vectors as input has almost random performance, as expected. Yet, Figure~\ref{fig:res} shows that both sets of $z$ vectors are useful for obtaining calibration parameters that generalize across conditions. Hence, we might hypothesize that whatever speaker information present in those vectors is being discarded when computing $\alpha$ and $\beta$ from the $z$ vectors using Equations (\ref{eq:alpha}) and (\ref{eq:beta}). This is an open question for future work.

\begin{figure}[!t]
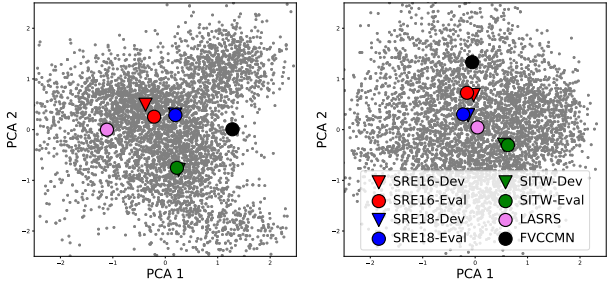

\centering
\begin{subfigure}{}
\includegraphics[width=0.48\columnwidth]{{{figs/metadata_analysis_old_model}}}
\end{subfigure}
\begin{subfigure}{}
\includegraphics[width=0.48\columnwidth]{{{figs/metadata_analysis_new_model}}}
\end{subfigure}
\vspace{-0.5cm}
\caption{Scatter plots of first and second PCA dimensions for CA-DPLDA (on the left) and AS-DPLDA (on the right) $z$ vectors, averaged per dataset. Grey dots correspond to a random sample of the PCA-projected values for the training data. The triangle markers that are hard to find are behind the corresponding circles.}
\label{fig:scatterz}
\end{figure}

We can also qualitatively analyze the nature of the $z$ vectors by plotting them in two dimensions after projection with principal component analysis, training this projection using the balanced training set used in the second stage of training.  Figure~\ref{fig:scatterz} shows the center of the projected $z$ vectors for each dataset, along with a sample of the individual training points, for reference. We can see that, in both cases, the centers for the development and evaluation sets from the same collection fall relatively close to each other. This seems to indicate that both $z$ vectors preserve information about the conditions of the samples, apparently in similar degrees. 

Interestingly, we can see that the FVCCMN center lies outside of the main mass of training data for both systems, more extremely for the CA-DPLDA case. This is expected given that, as explained before, this data is unlike any of the training data. This might explain why the calibration with DPLDA is not working well for this dataset: since, during training, the model has not seen enough $z$ values in the region where the FVCCMN $z$ values lie, its prediction of the calibration parameters for this data is not reliable. In the future we will look into detecting these cases where the model is doomed to fail, in a similar way as we did for TBC in \cite{Ferrer:aslp18}.

\subsection{Effect of initialization and number of epochs}
\label{sec:init}

\begin{figure}[!t]
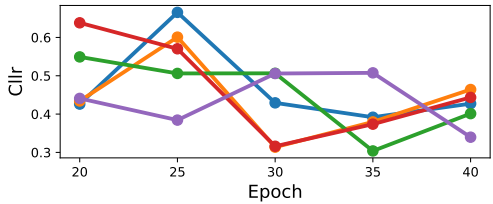

\centering
\includegraphics[width=0.8\columnwidth]{{{figs/performance_by_iter}}}
\vspace{-0.25cm}
\caption{Average actual Cllr on the development sets for the proposed method as a function the epoch for five initialization seeds.} 
\label{fig:perfbyiter}
\end{figure}

The results shown in Figure \ref{fig:res} correspond to the systems that lead to the best results over 10 different seeds and 5 different epochs (20, 25, 30, 35 and 40) on average over the development sets. This optimization turns out to be essential for good performance. Figure \ref{fig:perfbyiter} shows the average actual Cllr over the development  sets for 5 different seeds over different epochs for the AS-DPLDA model. We can see that different seeds and epochs can lead to very different results. In fact, even for the same seed, nearby epochs vary greatly in performance suggesting that the model is jumping from one local minimum to another, some of them having much better generalization performance than others. Note that this behavior is also observed for the originally proposed model that required external condition vectors. Lower learning rates and larger regularization values do not solve this problem. 

\begin{figure}[!t]
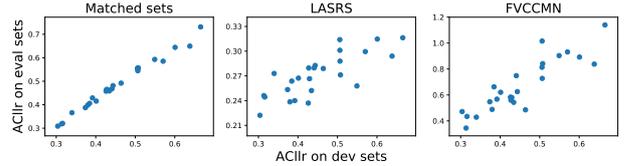

\centering
\includegraphics[width=1.0\columnwidth]{{{figs/generalization}}}
\vspace{-0.45cm}
\caption{Average actual Cllr on different subsets of the evaluation sets versus the development sets for different seeds and epochs. The evaluation sets are separated in three subsets: matched sets (SRE16-Eval, SRE18-Eval and SITW-Eval), LASRS and FVCCMN} 
\label{fig:perfbyiter_scatter}
\end{figure}

While it would be desirable for the performance to be more stable across seeds and epochs, the good news is that the best model selected using the development set is also a very good model on the evaluation sets. Figure \ref{fig:perfbyiter_scatter} shows the scatter plot of development versus evaluation performance for the same seeds and epochs as in Figure \ref{fig:perfbyiter}. We can see that on the sets that are matched to the three development sets, the correlation between development and evaluation performance is almost perfect. For the other two eval sets that do not have a corresponding development set, the correlation is not perfect, but we can still see that the best system for the development sets is, in both cases, a very good system for the eval set. Hence, the selection of the best model generalizes well in our experiments, even for cases for which no matching development set is available.

\begin{figure}
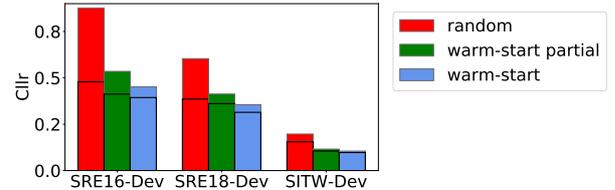

\centering
\includegraphics[width=1.0\linewidth]{{{figs/TABLE.initialization.ACLLR}}}
\vspace{-0.6cm}
\caption{Actual and minimum (black lines) CLLR for different initialization methods for the AS-DPLDA approach: all random, partial warm-start, and full warm-start (see text for details).}
\label{fig:initialization}
\end{figure}

Finally, Figure \ref{fig:initialization} shows the effect of the initialization method described in Section \ref{sec:dplda}, where all parameters in the model are initialized with some meaningful value, except the DimRed stage, which has no obvious default value. We call this method ``warm-start". Further, we wish to evaluate how important it is to initialize the extractor of $m$ vectors with the lower part of the LDA transform. Hence, we disable this initialization and replace it with random initialization, while all other parameters are initialized as in the warm-start approach. We call this method ``warm-start partial". Finally, we compare these two approaches with random initialization for all parameters. All random initializations are done using a normal distribution centered at 0.0 with standard deviation of 0.5. Of course, the distribution of the random initializers could be optimized, perhaps separately for each parameter, but this would be a very costly experiment. In all cases, the best model over 10 different seeds is selected. We can see that using the precomputed parameter values for initialization leads to significantly better results than random initialization and smaller but also consistently better results than using a random matrix to initialize the $m$ vector extractor instead of the lower part of the LDA transform. 

\section{Conclusions}
We presented a novel backend approach for speaker verification which consists of  a series of operations that mimic the standard PLDA-backend followed by calibration. The parameters of the model are learned jointly to optimize the overall speaker verification performance of the system, directly targeting the loss of interest in the speaker verification task. In order to achieve good generalization in terms of calibration performance across varying conditions, we introduced a side-information dependent calibration stage, where the side-information is learned jointly with the rest of the model.  We showed that the proposed approach improves performance over the standard PLDA backend on a wide variety of test conditions, leading to a robust backend that does not require specific development data for calibration.  

The proposed approach fails to give good calibration performance on only one of our test sets, which contains conditions that are not represented in the training data. We believe this is a case where the system should be able to reject the trials for being severely mismatched to the training data. We plan to pursue this research direction in the near future. Further, we plan to extend the approach to be able to score multiple enrollment trials and use additional external side-information like the signal's duration as input to the calibration model. Finally, the end goal is to integrate this backend in the embeddings extractor DNN for joint training.

\bibliographystyle{IEEEbib}
\bibliography{./all-short}

\end{document}